\documentclass[onecolumn,amssymb,fleqn,prb]{revtex4} %fleqn: make all eqns left alighed
\usepackage{epsfig,amssymb,amsmath,graphicx,subfigure}  % pdfpages and graphicx are added   hyperref(set hyperlink for content)
\usepackage{color}
\usepackage{hyperref}
\newcommand{\be}{\begin{eqnarray}}   %%\nonumber\\
\newcommand{\ee}{\end{eqnarray}}

\begin{document}

\title{The electric double layer structure around charged spherical interfaces}
\author{Zhenwei Yao, Mark J. Bowick and Xu Ma}
\affiliation{Department of Physics, Syracuse University, Syracuse,
New York 13244-1130, USA}

\begin{abstract}
We derive a formally simple approximate analytical solution to the
Poisson-Boltzmann equation for the spherical system via a
geometric mapping. Its regime of applicability in the parameter
space of the spherical radius and the surface potential is
determined, and its superiority over the linearized solution is
demonstrated.

\end{abstract}
%%\pacs{82.45.Gj}
\maketitle

Charged objects in electrolyte solutions are surrounded by
electric double layers (EDL) \cite{Ohshima}. One ionic layer is
due to a host of chemical interactions, and the second layer is
formed by the excess of oppositely charged ions in the solution,
screening the charged objects. The EDL structure is responsible
for the stability of colloidal dispersions \cite{VerweyNOverbeek}
and various electrostatic phenomena in biophysical systems
\cite{highsi_1}. The distribution of the screening potential in
EDL is characterized by the Poisson-Boltzmann (PB) equation. In
this letter, we will study the EDL structure around charged
spherical interfaces, which are ubiquitous in colloidal and
biophysical systems \cite{Colloid,Bio}.

Largely due to its nonlinear nature, analytical solutions to the
PB equation are available only for planar \cite{analytical_p} and
cylindrical \cite{analytical_cyl} systems. For a spherical system,
the analytical solution to the linearized PB equation is available
\cite{VerweyNOverbeek}. Despite a recently proposed analytical
series solution \cite{series_solution}, a formally simple
approximate solution is still in demand for studying analytical
problems that are based on the screening potential in EDL.
Numerical techniques\cite{num_Loeb} and the Debye-Huckel
linearized approximations have long been the only available basic
methods to solve the PB equation. Various perturbative solutions
have been proposed based on the planar solution
\cite{App_sphe_Plane}, or the linearized solution
\cite{App_sphe_DH}, as the zeroth order approximation.
Perturbative methods that start with the planar solution are
limited to the regime of large spherical radius, while those which
start with the linearized solution work in the weak potential
regime. The geometric construction of a formally simple
approximate analytical solution that can match both the planar and
the linearized solutions is one concern of this letter.

\begin{figure}[t]
  \centering
  \includegraphics[width=2in]{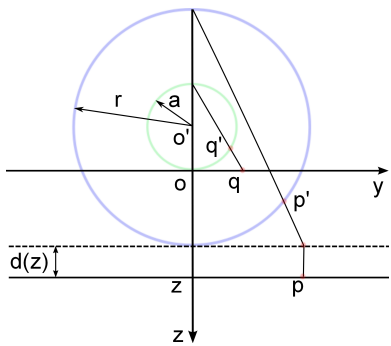}
  \caption{The mapping from the $z>0$ bulk space to the space
    outside a sphere with radius $a$ (represented by the blue circle)
    via. consecutive stereographic projections from a plane to a
    sphere. A deviation $d(z)$ is introduced to guarantee
    the equality of the potential on the blue spherical shell and
    the potential at the plane located at z.} \label{p_s}
\end{figure}

In q:-q symmetric electrolytes, the dimensionless
Poisson-Boltzmann equation is \cite{VerweyNOverbeek} \be \Delta
\psi=\sinh\psi ,\label{PB_dimensionless} \ee in the natural units
$k_B T/q$ and the Debye length $\kappa^{-1}= \sqrt{\epsilon k_B
T/(8 \pi n q^2)}$. $q$ is the absolute value of the charge of
ions. We propose an analytic scheme to yield an approximate
solution to the Poisson-Boltzmann equation in a spherical system
from the known analytical solution to the planar problem. The
planar and spherical systems are connected by a geometric mapping
defined in Fig. \ref{p_s}. Both the planar and spherical systems
are schematically plotted in Fig. \ref{p_s}. The planar system is
composed of a charged plate locating on the x-y plane with the
solution in the $z>0$ bulk space; and the corresponding spherical
system consists of a charged sphere of radius $a$ centered at the
origin, which is immersed in the solution in the $r>a$ bulk space.
The surface potential in both systems is denoted as $\psi_0$. The
potential in the planar system is known as $\psi(z)$, while that
around the corresponding spherical system is $\psi(r)$. These two
potentials can be related by a geometric mapping which is defined
as follows. By moving an arbitrary equipotential plane at $z$ in
the planar system by $d(z)$ followed by a stereographic projection
as shown in Fig. \ref{p_s}, the equipotential plane $\psi(z)$ in
the plate system is geometrically mapped to the equipotential
spherical shell $\psi(r(z))$ in the corresponding spherical
system. The charged plate on the x-y plane is mapped to the
spherical interface at $r=a$, so $d(z=0)=0$. The whole $z>0$ bulk
space in the planar system can be mapped to the bulk space outside
the spherical interface by repeating the mapping defined above for
all equipotential planes below the x-y plane.

The displacement field $d(z)$ is introduced to guarantee that \be
\psi(z)=\psi(r) \label{p_sr} \ee with $r=a+z-d(z)$. $d(z)$ encodes
all information of the potential about the spherical interface.
Geometrically $d(z)$ describes how the equipotential planes in the
planar system squeeze to form the equipotential spherical shells
in the corresponding spherical system. The problem for solving the
PB equation in a spherical system is now converted to solving for
the geometric deviation $d(z)$. The form of $d(z)$ can be found in
the weak potential regime where both $\psi_L(z)$ of a planar
system and $\psi_L(r)$ of a spherical system are known: \be
\psi_L(z)=\psi_0 e^{-z} ,\label{sol_plate_weak}\ee and \be
\psi_L(r)= \psi_0 \frac{a}{r} e^{a-r} \label{sol_s}, \ee where $a$
is the radius of the spherical interface and the subscript L
stands for linearized solution. Inserting Eq.
(\ref{sol_plate_weak}, \ref{sol_s}) into Eq. (\ref{p_sr}) leads to
$a/(a+z-d(z))= \exp(-d(z)) $, from which we have \be d(z)= a+z -
W(a e^{a+z}) ,\label{dz}\ee where $W(x)$ is the Lambert's W
function defined by $x=W(x)\exp(W(x))$ \cite{Wfunction_math}. It
is checked that $d(z\rightarrow 0)=0$ and $d(z) \rightarrow 0$ as
$a\rightarrow \infty$. And $d(z\rightarrow \infty)\sim \ln(z/a)$,
since asymptotically $W(x \rightarrow \infty)\sim \ln x - \ln (\ln
x)$ (Ref. [12]). The Lambert's W function is also found in other
physical systems, such as the fringe field of a capacitor and
Wien¡¯s displacement law in black body radiation
\cite{Wfunction_physics}. Eq. (\ref{dz}) shows that $d(z)$ is
independent of $\psi_0$ in the weak potential limit, since
$\psi_0$ appears as a prefactor in both $\psi(z)$ and $\psi(r)$ in
the weak potential limit as shown in Eq. (\ref{sol_plate_weak},
\ref{sol_s}).

\begin{figure}[t]
  \centering \subfigure[]{
    \includegraphics[width=1.585in]{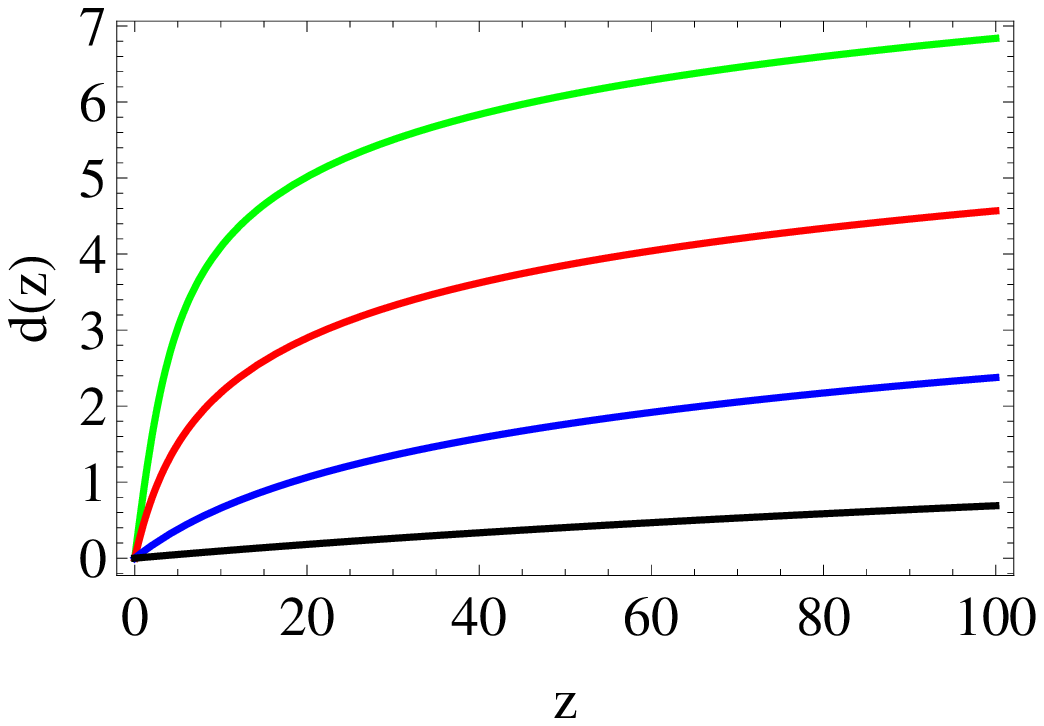}}
  \hspace{-0.01in} \subfigure[]{
    \includegraphics[width=1.66in]{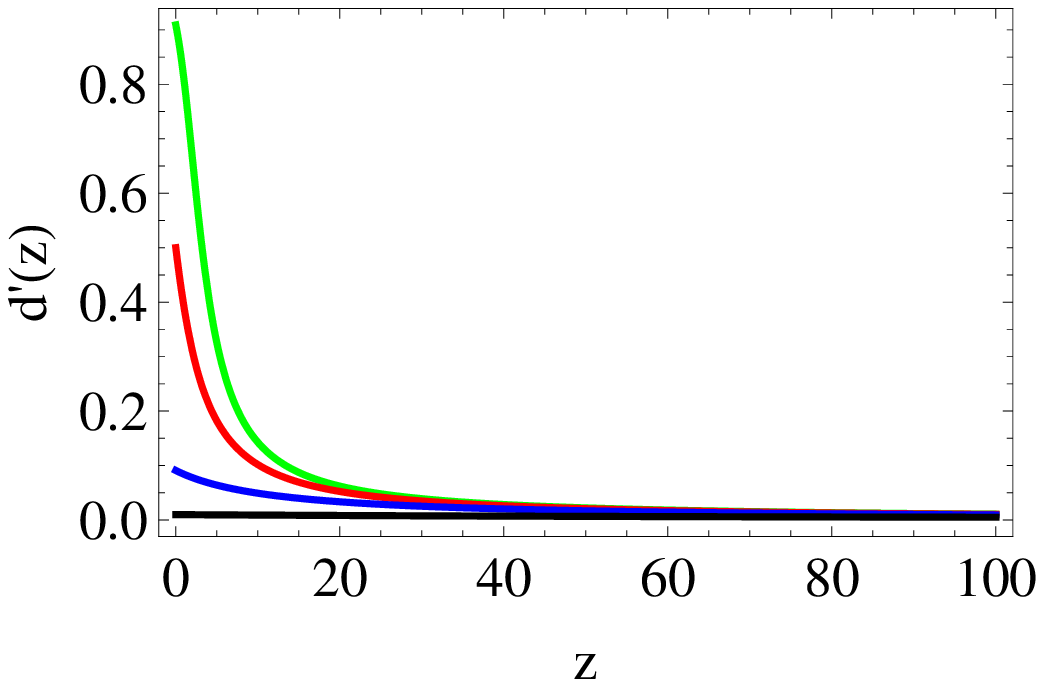}} % bb=16 0 380 275
  \caption{(a) The plot of the displacement field $d(z)$ for various
    spherical radii: $a=0.1$ (green), $a=1$ (red), $a=10$ (blue),
    $a=100$ (black). As $a\rightarrow \infty$, $d(z)$ is expected to
    vanish. (b)  The plot of the strain fields $d'(z)$ for the
    corresponding displacement fields $d(z)$ in (a). } \label{plotdz}
\end{figure}

The plot of $d(z)$ for various spherical radii is given in Fig.
\ref{plotdz} (a). The squeezing of equipotential surfaces near a
spherical interface with smaller radius is seen to be larger. It
is expected that the displacement vanishes for an infinitely large
spherical interface that approaches a plate. Fig. \ref{plotdz} (a)
also shows the behaviors of $d(z)$ in two regions -- steep slope
for small $z$ and much smaller slope for large $z$ where
$d(z\rightarrow \infty)\sim \ln(z/a)$. It gives the qualitative
picture of forming equipotential spherical shells from the
corresponding equipotential planar slices. Near the charged
spherical interface each slice moves more than the slice ahead of
it that is closer to the interface. In the region far away from
the interface, the squeezing is much more uniform. The relative
squeezing of equipotential surfaces is better demonstrated in
terms of the \textquotedblleft strain'' field $d'(z)$. The strain
field plotted in Fig. \ref{plotdz} (b)
 shows that the strain is
concentrated near the spherical interface and the strain
concentration is more significant near spherical interfaces of
smaller radii.

We suggest that the form of $d(z)$ for arbitrary $\psi_0$ be
approximated by Eq. (\ref{dz}) under the assumption of weak
dependence of $d(z)$ on $\psi_0$. This assumption is to be
substantiated later. We can then construct the analytical
approximate solution to the PB equation for a spherical system
from the known analytical solution to a planar system, which is
\be \psi=2\ln \frac{1+\gamma e^{-z}} {1-\gamma e^{-z}}
,\label{sol_p1}\ee where $\gamma=
(\exp(\psi_0/2)-1)/(\exp(\psi_0/2)+1)$. On the other hand,  $r(z)=
W(a \exp(a+z))$ and $W(a \exp(a+z)) \exp(W(a
  \exp(a+z))) = a \exp(a+z)$ yield \be z(r)= r-a+\ln(r/a) .\ee
The approximate solution denoted as $\psi_G(r)$ for the spherical
system is thus derived as \be \psi_G(r) =2 \ln [\frac{1+\gamma
\exp(-(r-a+\ln(r/a)))} {1-\gamma \exp(-(r-a+\ln(r/a)))}]
,\label{sol_G}\ee where the RHS is the potential in the
corresponding plate system with $z$ replaced by $z(r)$. Near a
spherical interface of large radius, i.e., $(r-a)/a<<1$ and
$a>>1$, the $\psi_G$ solution approaches the planar solution, as
required. In the region far away from the interface ($r>>a$), Eq.
(\ref{sol_G}) becomes $\psi_G(r)=4 \gamma a \exp(-(r-a))/r$, which
reduces to the linear spherical solution Eq. (\ref{sol_s}) in the
weak potential limit. Note that the $\psi_G$ solution may be
derived algebraically by a variable substitution $s=a/r
\exp(-(r-a))$ in Eq. (\ref{PB_dimensionless}) and more accurate
results can be obtained by perturbation analysis
\cite{White,White2}. In comparison to the algebraic method, the
derivation of the $\psi_G$ solution via the geometric mapping not
only reduces the complexity of algebraic calculations, but also
shows how the spherical geometry modifies the equipotential
surfaces of a planar system as encoded by the geometric deviation
$d(z)$. The relation between the $\psi_G$ solution and both the
linearized and planar solutions is also revealed in the geometric
derivation.

\begin{figure}[t]
  \centering \subfigure[]{
    \includegraphics[width=1.65in]{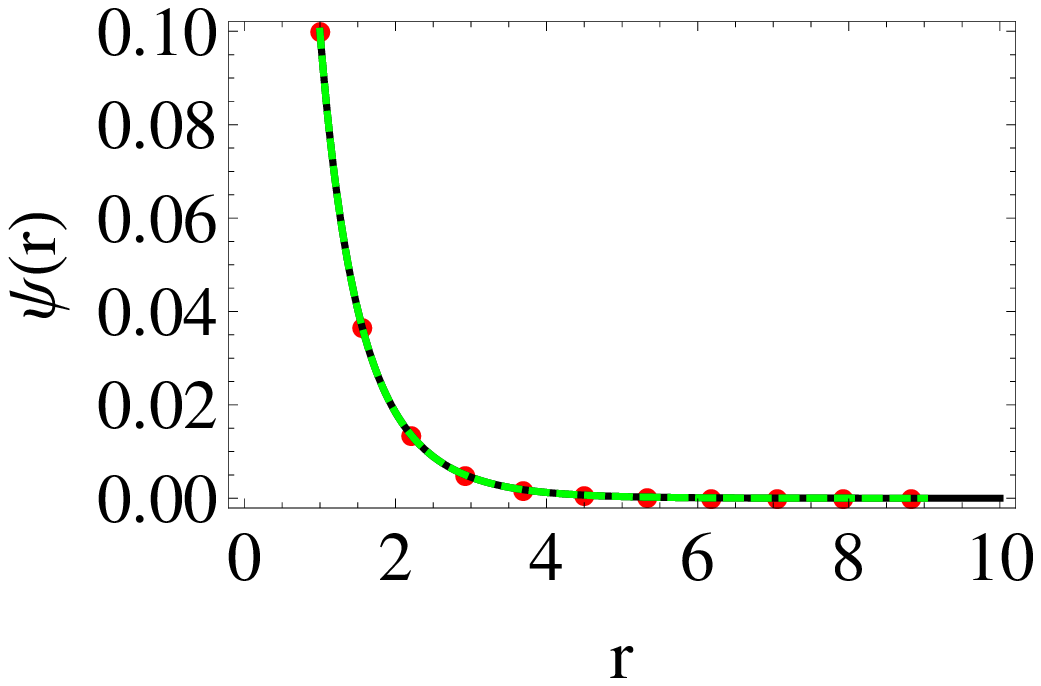}}
  \hspace{-0.01in} \subfigure[]{
    \includegraphics[width=1.6in]{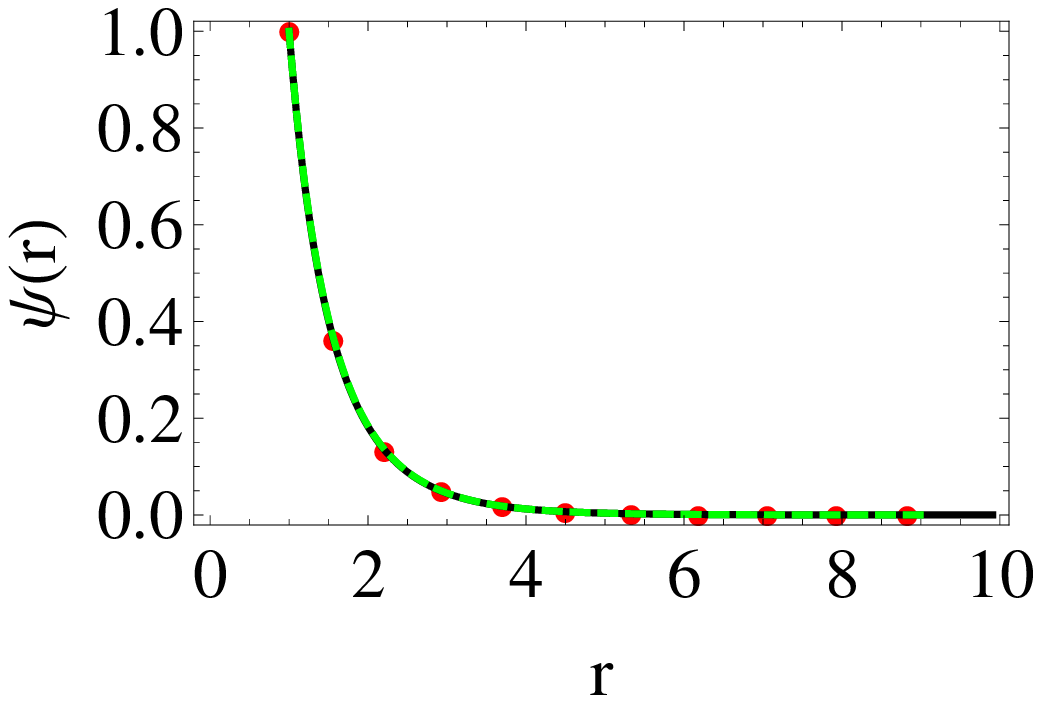}} % bb=16 0 380 275
  \hspace{-0.01in} \subfigure[]{
    \includegraphics[width=1.6in]{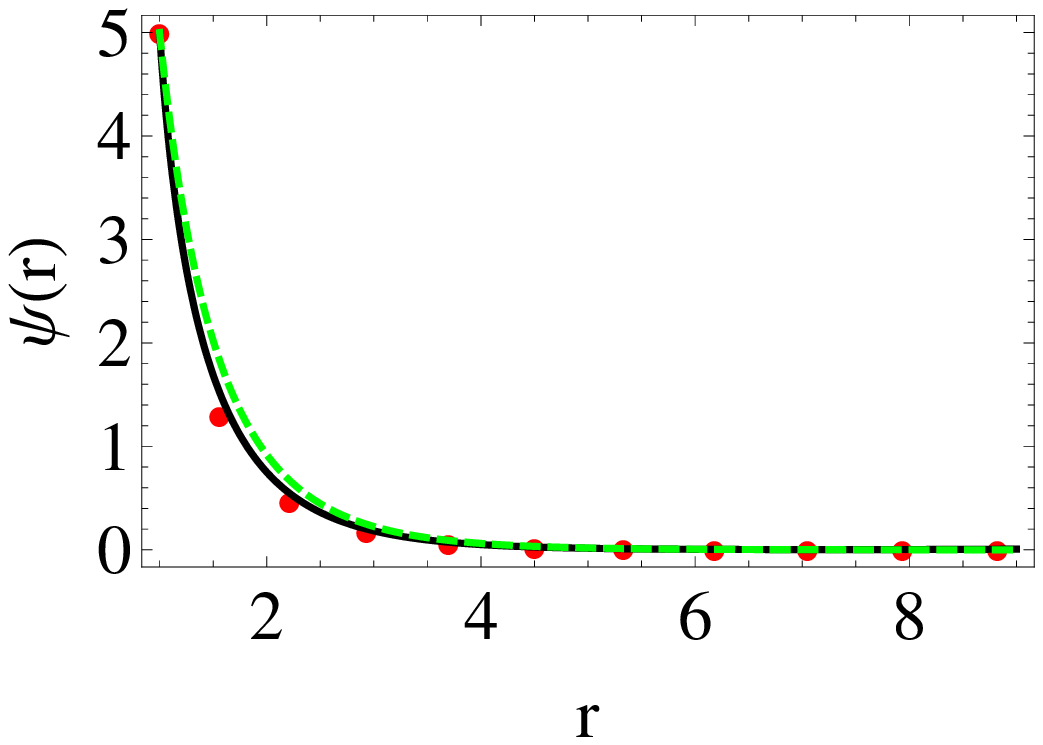}} % bb=16 0 380 275
  \hspace{-0.01in} \subfigure[]{
    \includegraphics[width=1.62in]{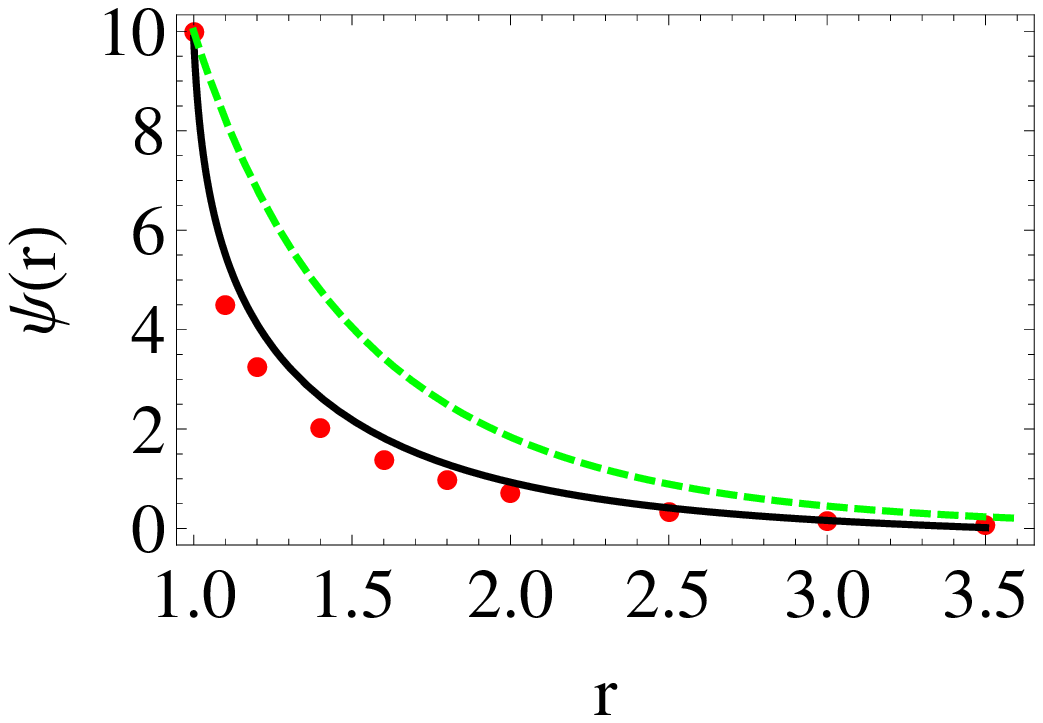}} % bb=16 0 380 275
  \caption{The potential around a charged spherical interface of unit
    radius in a q:q solution. The surface potentials in
    (a)-(d) are respectively $0.1, 1,5,10$ measured by $q/(k_B T)
    \equiv 1$. The curves give the numerical solution to the
    Poisson-Boltzmann equation (black), analytical solution for weak
    $\psi_0$ (green dashed) and the solution constructed by the our
    method (red dots). } \label{four}
\end{figure}

Eq. (\ref{sol_G}) is derived from the planar and the linearized
spherical solution, so at the very least it is expected to work
for either $a>>1$ or $\psi_0<<1$. It is therefore superior to the
linear solution, which only works in the weak potential limit. The
region of validity of the $\psi_G$ solution can be derived
algebraically. By introducing $x=1/r$, the PB equation for a
spherical system becomes \cite{same_equation}
\begin{equation} \label{eq:1}
C(x)\frac{\partial^{2}}{\partial{}x^{2}}\psi=\sinh\psi ,
\end{equation}
where $C(x)=x^{4}$. Inspired by the functional form of the planar
and the linearized spherical solutions Eqs.(\ref{sol_plate_weak},
\ref{sol_s}), we use $\psi(x)=2\ln ((1+g(x))/(1-g(x)))$ as a trial
solution. Depending on the sign of $\psi(x)$,
$g(x)=\pm{}\exp(-f(x))$ and $g(x)\in(-1,1)$ corresponding to $\psi
\in (-\infty,\infty)$. Inserting the ansatz into Eq. (\ref{eq:1})
yields
\begin{equation}
\label{eq:2} g(x) [C(x)(f'^{2}-f'')-1] +g^{3}
[C(x)(f'^{2}+f'')-1]=0.
\end{equation}
For $|g|<<1$, by dropping the $g^3$ term, the solution to Eq.
(\ref{eq:2}) is $f(x)=1/x-\ln{x} + c_1$, with an integration
constant $c_1$. Inserting $f(x)$ into the ansatz $\psi$ yields the
$\psi_G$ solution. An alternative condition for dropping the $g^3$
term in Eq. (\ref{eq:2}) is $C(x)(f'^{2}+f'')-1<<1$, which is
equivalent to $x=1/r<<1$ by inserting the expressions for $f(x)$
and $C(x)$. Therefore, for either $|g|<<1$ or $x=1/r<<1$, the
solution to Eq. (\ref{eq:2}) coincides with the $\psi_G$ solution.
Note that $|g|<<1$ is equivalent to the weak potential limit, and
$x=1/r<<1$ holds for $a>>1$ since $r>a$. An important case falls
in this region of validity of the $\psi_G$ solution. Consider
colloids of size $R$ in a solution of ion strength $I$ (in mol/L).
The Debye length is
$\kappa^{-1}(\textrm{nm})={0.304}/{\sqrt{I(\textrm{mol/L})} } $
which is at the order of nm for $I \sim 1\ \textrm{mol/L}$
\cite{Intermolecular}. For $R\sim \ \mu \textrm{m}$,
$a=R/\kappa^{-1} >> 1$. Therefore, the approximate analytical
$\psi_G$ solution is suitable for typical colloidal dispersions.
In comparison to the series solution \cite{series_solution}, the
formal simplicity of the $\psi_G$ solution enables further
analytical study of the electrostatics of colloidal systems.

The $\psi_G$ solution turns out to have a larger region of
validity. Fig. \ref{four} shows comparisons of the $\psi_G$
solution (red dots), the linearized solution (green dashed) and
the numerical solution (black) to the PB equation for a spherical
system for different potentials. For weak potential ($\psi_0=0.1
(a), 1 (b)$), the three solutions fall on the same curve. The
linearized solution works well at least up to $\psi_0=5$ without
qualitatively deviating from the numerical solution. So the
linearized theory applies for moderate values of surface potential
\cite{Bier}. As $\psi_0$ exceeds 5, the linearized solution starts
to deviate from the numerical solution, while $\psi_G$ conforms to
the numerical solution up to $\psi_0=10$, where the linearized
solution deviates significantly the numerical solution. This
indicates that the dependence of $d(z)$ on $\psi_0$ is weak for
$a=1$ up to $\psi_0=10$.

\begin{figure}[h]
  \centering
  \includegraphics[width=2in]{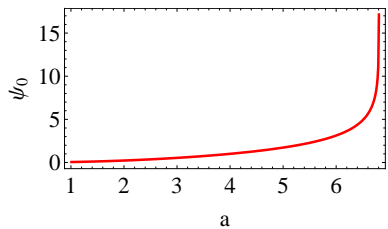}
  \caption{The $\psi_G$ solution is applicable below the red curve in the parameter space $\{a,
\psi_0 \}$. On the red curve,
$\delta=\epsilon=10^{-3}$.}\label{deviation}
\end{figure}

\begin{figure}[h]
  \centering \subfigure[]{
    \includegraphics[width=1.6in]{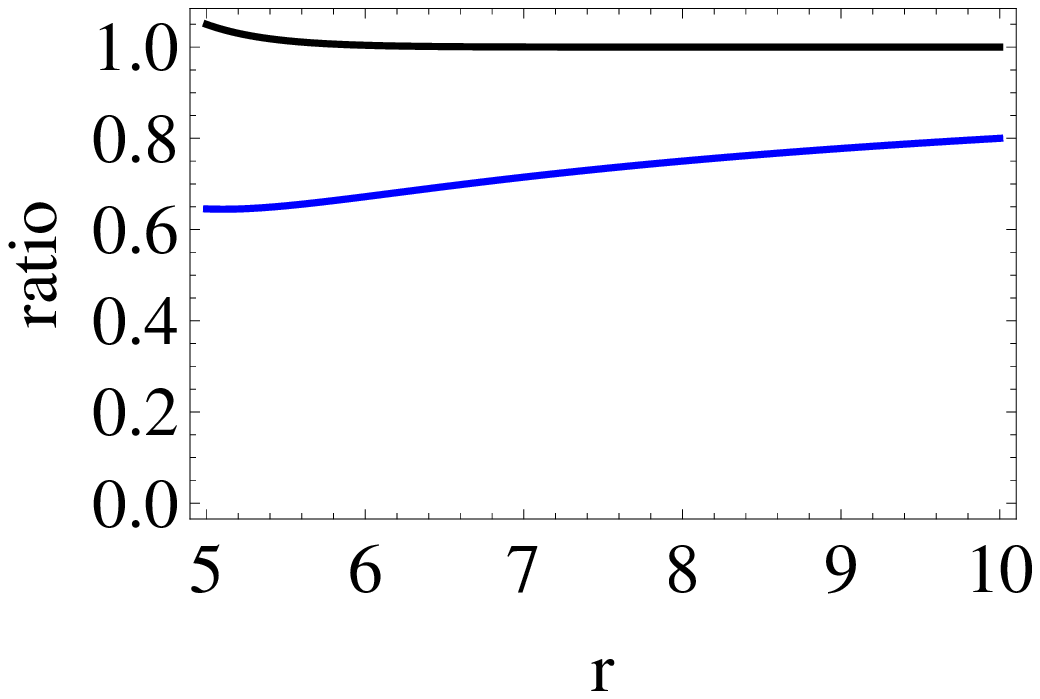}}
  \hspace{-0.01in} \subfigure[]{
    \includegraphics[width=1.6in]{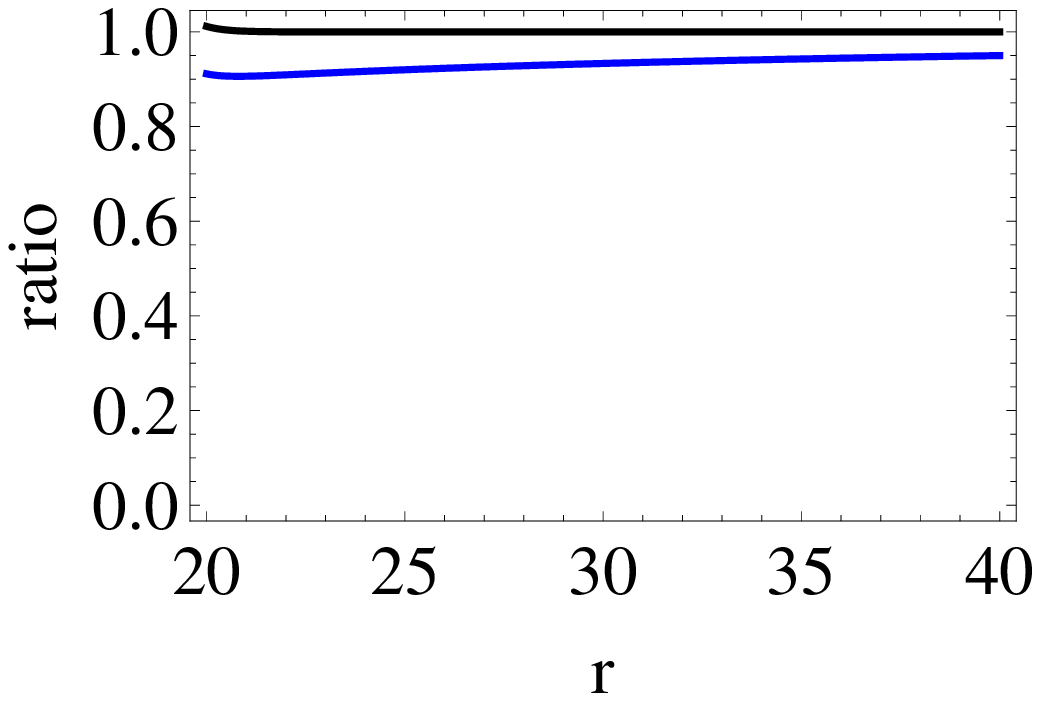}} % bb=16 0 380 275
  \caption{The comparison of the $\psi_G$ solution (black) and the planar solution
  (blue) in terms of the ratio of the LHS to RHS of Eq. (\ref{PB_dimensionless}).  The radii of
    the spherical interface are $a=5 (a)$ and $20 (b)$. $\psi_0=1$. } \label{big_psi_plot}
\end{figure}

The quality of the $\psi_G$ solution  can be systematically
studied by defining a ratio $\delta=\max_{r}
\{|(\Delta\psi-\sinh\psi)/\sinh\psi|\}$. The smaller the ratio
$\delta$ is, the better the solution is. For a given precision
$\epsilon=10^{-3}$, the applicable region of the $\psi_G$ solution
is found to be below the red curve in the parameter space $\{a,
\psi_0 \}$, as shown in Fig.  \ref{deviation}. For $a \gtrsim 7$,
the $\psi_G$ solution applies even for large potentials. There
exists, however, a cut-off value for the surface potential. High
potential, or equivalently low temperature, may lead to
correlation of counter-ions near the charged interface that is
ignored in the mean field PB equation \cite{highsi_1}. In
addition, high potential leads to high concentration of ions so
that the finite dimension of ions must be taken into consideration
\cite{VerweyNOverbeek}. The advantage of the $\psi_G$ solution
over the planar solution is shown explicitly in Fig.
\ref{big_psi_plot}. The $\psi_G$ solution works better than the
planar solution even for large spherical radius.

In conclusion, we have studied the EDL structure around charged
spherical interfaces by analysis of the Poisson-Boltzmann
equation. Despite the point charge assumption of electrolyte ions
and the neglect of ion-ion corrections, the PB equation generally
works well especially for problems of electrostatic interaction of
colloidal particles \cite{Ohshima}. In this paper, we have derived
an approximate analytical solution to the Poisson-Boltzmann
equation for the spherical system by a geometric mapping. The
formal simplicity of the $\psi_G$ solution enables further
analytical study of spherical systems. The regime of applicability
includes not only the weak potential regime where the linearized
solution also works well, but also the regime of large spherical
radius. Typical colloidal dispersions with the size of colloids
much bigger than the Debye length fall in the latter regime.

\acknowledgments We thank Shiliyang Xu for discussions. This work
was supported by the National Science Foundation grant DMR-0808812
 and by funds from Syracuse University.


\begin{thebibliography}{0}
\bibitem{same_equation}
{The PB equation can be written in a unified form: $C(x)
\psi''(x)=\sinh \psi(x)$ in the planar ($C(x)=1$ with $x=z$),
cylindrical ($C(x)=\exp(-2x)$ with $x=\ln{r}$) and spherical
($C(x)=x^{4}$ with $x=1/r$) cases. }
\end{thebibliography}
\end{document}